\documentclass[twocolumn,prb,superscriptaddress,showpacs]{revtex4}
\usepackage[T1]{fontenc}
\usepackage{amsmath,amsbsy,amssymb,graphicx}
\usepackage{times}
\let\mathbf=\boldsymbol

\begin{document}

\title{ Spin-filtered and Spatially Distinguishable Crossed Andreev Reflection in a Silicene-Superconductor Junction}
\author{Kangkang Li}
\affiliation{Institute of Physics, Chinese Academy of Sciences, Beijing 100190, China}
\author{Yan-Yang Zhang}
\email{yanyang@semi.ac.cn}
\affiliation{SKLSM, Institute of Semiconductors, Chinese Academy of Sciences, P.O. Box 912, Beijing 100083, China}
\affiliation{Synergetic Innovation Center of Quantum Information and Quantum Physics,
University of Science and Technology of China, Hefei, Anhui 230026, China}

\begin{abstract}
We theoretically investigate the quantum transports in a junction between a superconductor and a silicene nanoribbon, under the effect of a magnetic exchange field.
We find that for a narrow nanoribbon of silicene, remarkable crossed Andreev reflection (with a fraction $>50\%$) can be induced in the energy window of the elastic cotunneling,
by destroying some symmetries of the system. Since the energy responses of electrons
to the exchange field are opposite for opposite spins,
these transport channels can be well spin polarized. Moreover, due to the helicity conservation of the topological edge states, the Andreev reflection, the crossed Andreev reflection and the elastic cotunneling are spatially separated in three different locations of the device, making them experimentally distinguishable. This crossed Andreev reflection is a nonlocal quantum interference between opposite edges through evanescent modes. If two superconducting leads with different phases are connected to two edges of the silicene nanoribbon, the crossed Andreev reflection can present Josephson type oscillations, with a maximal fraction $\sim 100\%$.
\end{abstract}

\maketitle

\section{Introduction}
The quantum entanglement between microscopic particles
attracts intense interests for its fundamental importance and
potential applications in
quantum information technology\cite{PR,GBL}.
Although entangled photons have been realized in experiments\cite{AA}, it is still a challenge to realize entangled electrons in condensed matter
systems. A superconductor (SC) is a condensate of Cooper pairs, composed of two
electrons with momenta and spins entangled\cite{PR,GB}. The crossed Andreev refection (CAR) is the Cooper pair forming from two interfaces at a distance less than the superconducting coherence length $\xi$. Therefore the inverse crossed Andreev reflection
is proposed to be a natural method to produce
entangled electrons, where a Cooper-pair in the superconductor are
split into two electrons in two spatially apart leads\cite{PR,GBL}.

\begin{figure}[t]
\centerline{\includegraphics[width=0.45\textwidth]{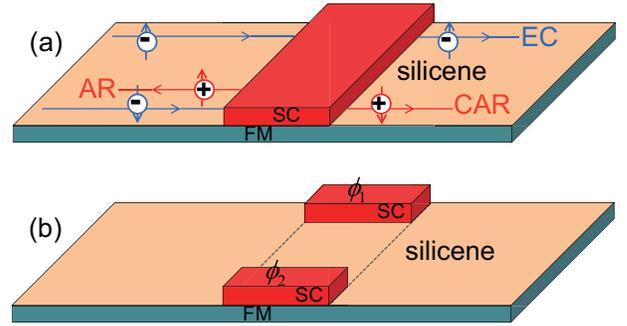}}
\caption{(Color online) Schematic of the proposed experimental setups. After the silicene nanoribbon is deposited on a
FM substrate, (a) the central region is entirely covered by a single SC lead, or (b) two edges of the central region are covered by two SC leads with phase $\phi_i$. In (a), we also illustrate the channel configuration at energy $E_0$ marked in Fig. \ref{FigspectrumToCAR} (g) and (h), where up and down arrows denote spin direction, and plus and minus signs represent
hole and electron edge states.}
\label{Figschematic}
\end{figure}

Although the CAR has already been experimentally observed in SC-metal\cite{DB,SR,PCZ}, or even SC-graphene \cite{GrapheneKim} heterostructures, usually there are
other coexisting and competing channels, e.g., the Andreev reflection (AR), the elastic cotunneling (EC) and the normal reflection (NR). The coexistence and mixture of these transport channels make it difficult to distinguish the CAR alone in the experiments.
Hence, there were many attempts
to enhance the CAR fraction\cite{GD,JN,MV,JC,Melin2008,JL,Linder2014,BenjaminEPL,YSAng2016,RMALY,Yeyati,JW2012,JW2015,WC}
or to separate the CAR channel from others in space\cite{RWR}.
The spatial separation of helical edge states in the quantum spin Hall (QSH)\cite{KM,Bernevig2006} systems makes it a good candidate for channel separation. As a result of the helical configurations of electron and hole edge states, the AR and EC are scattering processes within the same edge, while CAR and NR are those between opposite edges\cite{PA,RWR}, as illustrated in Fig. \ref{Figschematic}(a).
By bulk mediated transports, the CAR was proposed to be spatially separated from EC, but its fraction is small\cite{RWR}.
In this method, spin polarized currents were necessary for injecting, otherwise the CAR channel is inevitably mixed with the EC channel.

In this work, we investigate quantum transports in a silicene(S)-SC junction
under a ferromagnetic (FM) exchange field.
Silicene is a monolayer of silicone with a buckled honeycomb lattice structure, which has been fabricated recently\cite{SiliceneExperiment}.
Silicene is predicted to be in a QSH state with a small bulk gap\cite{KM,CCL}.
The device we investigate is illustrated in Fig. \ref{Figschematic}.
A silicene nanoribbon with zigzag edges is deposited on a FM substrate.
Then, in the central region, it is either entirely covered by an $s$-wave SC lead on the top [Fig. \ref{Figschematic} (a)], or covered by
two independent $s$-wave SC leads just on the top of two edges [Fig. \ref{Figschematic} (b)].
The length of the central region
$L$ is measured in units of lattice constant, and the width of the ribbon $W$ is measured as the number of silicon sites of the armchair chain traverse the sample.

The bulk gap of silicene is predicted to be only the order of meV\cite{CCL}, similar with the magnitude of the superconducting gap in conventional $s$-wave SCs. We find that, in this S-SC-S junction, this similarity of energy scales (QSH gap and SC gap) leads to interesting quantum effects, when the device width is comparable to associated characterized length scales (i.e., edge state penetration depth and SC coherence length).
Firstly, under the FM exchange field,
the original energy windows for AR plateaus are shifted in opposite directions for opposite spins.
Secondly, in the case of a single SC lead [Fig. \ref{Figschematic} (a)], when the onsite energy is turned on, some fraction (which can exceed 50\%) of the EC channel will turn into CAR.
Furthermore, in the case of two edge SC leads [Fig. \ref{Figschematic} (b)], the CAR can display Josephson-type periodic oscillations with the SC phase difference, and the CAR magnitude can be nearly $100\%$. Under the effect of the FM exchange field, the energy windows for the remarkable AR and CAR are spin dependent, so they can be spin filtered at a definite energy.
Most significantly, three transport channels, AR, EC and CAR can be engineered to
output at different locations of the device, as illustrated in Fig. \ref{Figschematic} (a),
which can be experimentally distinguishable by \emph{in situ} techniques\cite{EdgeExperiment,MajoranaExperiment}.

\begin{figure*}[t]
\centerline{\includegraphics[width=0.8\textwidth]{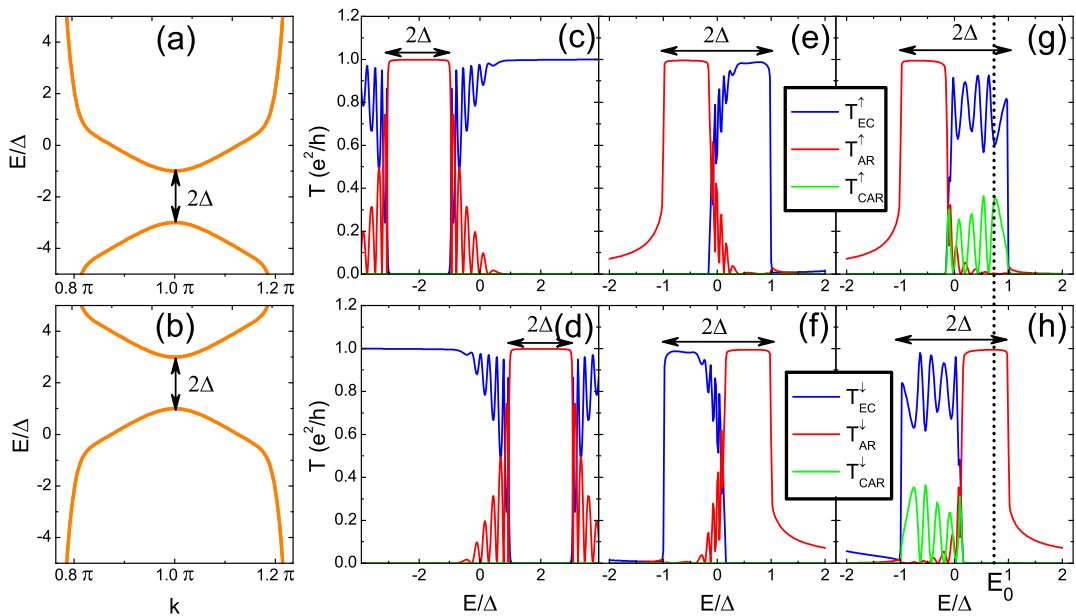}}
\caption{(Color online) (a) and (b) are BdG spectrums of the silicene ribbon in the ``pair method''
for the spin up (down) component respectively.
(c) and (d) are the spin resolved transmission coefficients calculated from the ``pair method''.
(e) to (h) are those calculated from the ``lead method'', with zero $U_S$ [(e) and (f)] and nozero $U_S$ [(g) and (h)]
respectively. The upper (lower) row corresponds the spin up (down) component.
The parameters are
$L=40$, $W=22$, $M=2$ meV and $U_S=1.5$ meV is only turned on in (g), (h).
}
\label{FigspectrumToCAR}
\end{figure*}

\section{Model and Methods}
The electronic states of silicene on a FM substrate can be described by the Hamiltonian\cite{CCL,ME,Yokoyama2013}:
\begin{align}
H_{\mathrm{S}}& =-t\sum_{\left\langle i,j\right\rangle \alpha }c_{i\alpha }^{\dagger
}c_{j\alpha }+i\frac{\lambda _{\text{SO}}}{3\sqrt{3}}\sum_{\left\langle
\!\left\langle i,j\right\rangle \!\right\rangle \alpha \beta }\nu
_{ij}c_{i\alpha }^{\dagger }\sigma _{\alpha \beta }^{z}c_{j\beta }  \notag \\
& +U_{S}\sum_{i\alpha}c_{i\alpha }^{\dagger}c_{i\alpha }
-M\sum_{i\alpha }c_{i\alpha }^{\dagger }\sigma _{z}c_{i\alpha },
\label{BasicHamil}
\end{align}%
where $c_{i\alpha }^{\dagger }$
($c_{i\alpha }$) is the creation (annihilation) operator of an electron on site $i$ with
spin $\alpha$; $\sigma_z$ is the Pauli matrix of spin and $\langle i,j\rangle
/\langle\langle i,j\rangle\rangle$ run over all the nearest/next-nearest
neighbor sites. The first term denotes the nearest-neighbor hopping with the transfer
energy $t=1.6$ eV. The second term is the intrinsic spin-orbit coupling (SOC) with
$\lambda _{\text{SO}}=3.9$ meV, and $\nu_{ij}=\pm1$ when the next-nearest-neighbor
hopping is anticlockwise (clockwise) with respect to the $+z$ axis. The third
term represents onsite energy $U_S$ in the central region which can be controlled by
a gate voltage\cite{JW2012,JW2015}. The last term represents the exchange field with strength $M$
arising from the ferromagnet substrate.\cite{JL,JW2012,JW2015}

The effect of the superconducting lead can be included in two different ways.
The first way (the ``pair method'') is to add a superconducting pair term $H_{\mathrm{pair}}=\sum_{i}\Delta c^{\dagger}_{i\uparrow}c^{\dagger}_{i\downarrow}+\mathrm{H. c.}$ to the bare Hamiltonian (\ref{BasicHamil}),
where $\Delta$ is the superconducting gap parameter\cite{RWR,WC}. This closed form $H_{\mathrm{S}}+H_{\mathrm{pair}}$ only contains the electronic degree of freedom in silicene,
and is convenient for analytical treatments\cite{RWR,WC}.
However it is essentially describing an \emph{isolated} superconducting silicene.
A more realistic way (the ``lead method'') is to couple the silicene to a semi-infinite SC lead, utilizing the standard method of Green's function\cite{QFS,QFSJPCM,YingTaoSR}.
See Appendix A for more details.
In most of our calculations, we adopt a typical experimentally accessible value $\Delta=1$ meV.
Compared to the effective Hamiltonian methods\cite{JC,Linder2014,YSAng2016}, such tight-binding simulations treat possible bulk and edge states ``naturally'' and on an equal footing. Furthermore, the effects of disorder and size confining can also be investigated.

With the given model parameters, the Hamiltonian (\ref{BasicHamil}) is in the QSH state with a bulk gap $2\lambda_{\mathrm{SO}}=7.8$ meV,
where spin degenerate edge states live.
The presence of SC pair $\Delta$ will open a SC gap $2\Delta$ around the Fermi energy.
Moreover, a finite exchange field $M$ will split the edge states towards negative (positive) energy for the spin up (down) component respectively, as illustrated in Fig. \ref{FigspectrumToCAR} (a) and (b)\cite{JW2012}. This energy shift is important for spin-filter mechanism in the following\cite{JL,JW2012,JW2015}.

The transmission coefficients (in units of $\frac{e^2}{h}$ throughout this manuscript)
$T_{\mathrm{EC}}$, $T_{\mathrm{AR}}$ and $T_{\mathrm{CAR}}$ through the central region can be expressed
in terms of the standard method of Green's functions\cite{QFSJPCM,SGC}, which are explicitly shown in Appendix B.
All the above Hamiltonians do not contain spin flip terms.
Consequently, possible nonzero $T_{\mathrm{EC}}$ must be between electron states with the same spin,
while possible nonzero $T_{\mathrm{AR}}$ and $T_{\mathrm{CAR}}$ must be between electron and hole states with opposite spins. Moreover, spin resolved versions of all these transmission coefficients, $T^{\sigma}_{\mathrm{AR}}$, $T^{\sigma}_{\mathrm{CAR}}$ and $T^{\sigma}_{\mathrm{EC}}$ are also well defined according to the spin $\sigma=\uparrow, \downarrow$ of the injecting electron state.

\section{The Rise of the CAR}

We first focus on the case of a single SC lead illustrated in Fig. \ref{Figschematic} (a), and the case of two SC lead will be discussed in Section \ref{SectionDisorderJosephson}. For a perfect QSH system without the exchange field, it has been theoretically predicted\cite{QFS,PA} and experimentally reported\cite{RRDu2012} that there is an AR plateau of $T_{\mathrm{AR}}=2$ in the SC gap, contributing from both spin components. Outside the SC gap, there are EC plateaus\cite{WC,PA}. In Fig. \ref{FigspectrumToCAR} (c) and (d), we plot the spin resolved transmission coefficients
of a \emph{narrow} nanoribbon
at a finite exchange field $M=2$ meV by adopting the simulation method of ``pair method''.
Notice that now the AR plateau window (with a width $2\Delta$) for spin up (down) component is shifted by $\mp M$ along the energy axis\cite{JW2012}.
This $M$ induced energy shift of transmission windows can be easily understood from the energy spectrum of the
Bogoliubov-de Gennes (BdG) equation of the ``pair method''(See Appendix C), as plotted in Fig. \ref{FigspectrumToCAR} (a) and (b).
Outside the exchange field shifted superconducting gaps,
$T^{\sigma}_{\mathrm{AR}}$ decays to zero soon with rapid resonant oscillations, while $T^{\sigma}_{\mathrm{EC}}\simeq 1-T^{\sigma}_{\mathrm{AR}}$ develops into a perfect plateau\cite{WC,PA}.

Results simulated by ``lead method'' are shown in Fig. \ref{FigspectrumToCAR} (e) and (f).
Compared to the ``pair method'', the main difference is that, the above pictures of AR and EC plateaus and their energy shifts are only valid within the energy window $(-\Delta,\Delta)$,
which is pinned by the semi-infinite SC lead as a
bath. Outside this energy window of the SC ground state, since the electrons do not have to be bounded as Cooper pairs, the AR can hardly happen; Similarly, now the injecting electron will be easier to be absorbed into the SC lead (which was absent in the ``pair method''), rather than transmitted to the right lead, so that the EC vanishes. We find that, by using the ``pair method'', the superconducting gap will drift unlimitedly with $M$, which is less realistic for the system we study here. Hereafter, we will only present simulation results from more physical ``lead method'',
which is more closely related to the experimental setup.

So far, the onsite energy $U_S$ of the central region has been set to zero, and
the CAR is zero. Once $U_S$ is turned on,
as shown in Fig.\ref{FigspectrumToCAR} (g) and (h),
the perfect AR plateaus are left untouched,
but some fraction of the ECs (blue lines) has been converted into CARs (green lines). Because of the finite $M$, this CAR is also spin polarized.
For example, let us fix at a definite energy $E_0$ marked by the vertical dashed line in Fig.\ref{FigspectrumToCAR} (g) and (h). The corresponding configuration of transport channels are illustrated in Fig. \ref{Figschematic} (a). An injecting spin down electron from the left lead (must be along the lower edge) will result in an almost perfect local AR back to the left lead, and no EC or CAR can happen.
On the other hand, an injecting spin up electron (must be along the upper edge) will exhibit EC or CAR to the right lead, but with negligible AR. Notice that, besides the spin filter, now we have realized a complete spatial separation of the CAR $T_{\mathrm{CAR}}^{\uparrow}$ [green line in Fig.\ref{FigspectrumToCAR} (g)] from other channels at energy $E_0$:
The only coexisting output channel in the right lead
is the EC [blue line in Fig.\ref{FigspectrumToCAR} (g)], but it is flowing along the \emph{opposite} edge. Another nonzero transport channel at this energy is the AR [red line in Fig.\ref{FigspectrumToCAR} (h)], but it is in the left lead. This spatially separated CAR is the first important finding in this manuscript.

In Fig.\ref{FigspectrumToCAR} (g) and (h),
$T_{\mathrm{CAR}}\simeq 1-T_{\mathrm{EC}}$ oscillates with the energy, which reflects the resonant competition between the topological edge states and the superconducting order, as explained in the following.
We have seen that, outside the AR plateau,
the robust edge states tend to transport the injecting electron through the sample [i.e., EC, blue lines in Fig.\ref{FigspectrumToCAR} (c), (d), (e) and (f)] without a local Cooper pair forming near the left boundary.
However, during its journey towards the right lead, this electron always has the probability of forming a Cooper pair with another electron (with opposite spin) which can only come from the opposite edge. We have checked that for this system, the width $W \sim 20$ is not small enough to open a subgap from the direct overlap of states on opposite edges\cite{BZ}, as required by the mechanism proposed in Ref. \cite{WC}.
Nevertheless, under the effect of the superconducting pair $\Delta$, the evanescent nature of the wavefunctions along the transverse direction is sufficient to mediate a Cooper pair forming\cite{WC,RWR}, as long as the width is not much larger than the superconducting coherence length $\xi$. Notice that both competing effects (intra-edge forward transport and inter-edge pairing)
are nonlocal in nature, which lead to a remarkable quantum interference throughout the central region, and finally result in the resonant competition between the EC and the CAR reaching the right lead.

The reason for vanishing CAR process at $U_S=0$ is due to the opposite pseudoparities between the
quantum channels in the leads and the central region which prohibits the scattering\cite{JW2012,pseudoparity1,pseudoparity3,pseudoparity2}.
The existence of this pseudoparity stems from the presence of the mirror reflection symmetry and
the particle-hole symmetry of the bare Hamiltonian, as well as the bipartite lattice structure of the lattice\cite{pseudoparity3,pseudoparity2}.
Finite $U_S$ breaks the particle-hole symmetry respect to the Fermi energy, and
relaxes the rigorous orthogonality between the scattering states, thus produces the
possibility of nonzero $T^{\sigma}_{\mathrm{CAR}}$. We have checked (but not shown here) that another way of breaking
the particle-hole symmetry, including a small real-valued next nearest neighbor hopping $t_2$
can also induce nonzero CAR.
Later, we will see other examples of producing CAR by breaking the mirror reflection symmetry in Section \ref{SectionDisorderJosephson}.

\section{The Effects of Varying $U_S$ and Sample Size}

\begin{figure*}[t]
\centerline{\includegraphics[width=0.95\textwidth]{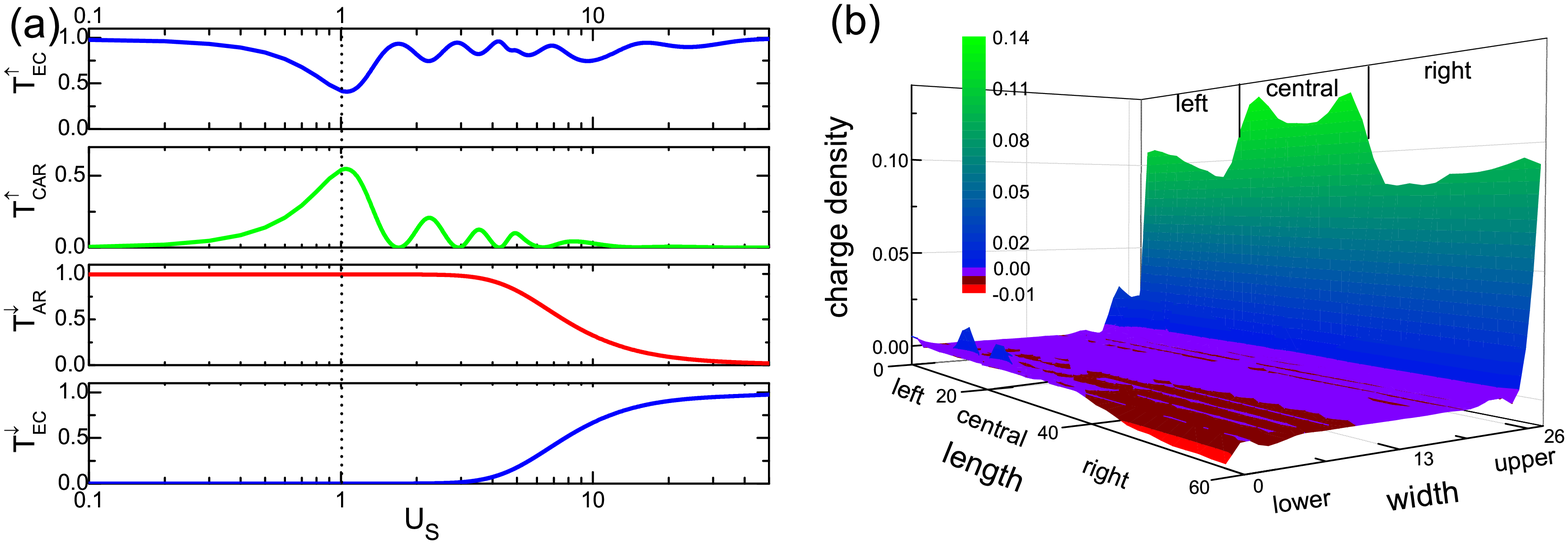}}
\caption{(Color online) (a) The transmission coefficients as
functions of the onsite energy $U_S$ of the central region. Other parameters are
$L=20$, $W=26$, $M=1.7$ meV and $E=0.66$ meV.
Notice that the $U_S$ axis is logarithmic.
(b) The real space charge density distribution in the device
with $U_S=1$ meV, corresponding to the dashed line in (a), where the optimal CAR occurs.
Some part of the left and right leads (marked as ``left'' and ``right'') are included in (b). Positive and negative numbers correspond to
electron and hole states respectively.}
\label{Figchargedensity}
\end{figure*}

In order to have a deeper insight on the role of $U_S$, in Fig. \ref{Figchargedensity} (a),
we show the dependence of transmission coefficients on the onsite energy $U_S$ at a definite energy $E$.
In the large $U_{S}$ limit, the central region has been doped into the bulk band,
and the dominating channel will be the EC for both spin components.
For the spin down electron, this leads to the decrease of AR,
whose perfectness was protected by the edge states.
This is why the AR plateau starts to collapse when $U_S$ touches the gap edge at $\lambda_{\mathrm{SO}}=3.9$ meV.
On the other hand, the behavior for spin up electron is more interesting for the oscillating patterns of $T_{\mathrm{EC}}$ and $T_{\mathrm{CAR}}$, which
also reflects the coherent competition between the nonlocal edge transport and the nonlocal superconducting pairing.
The highest peak of $T_{\mathrm{CAR}}$ reaches a maximum value $T_{\mathrm{CAR}}=0.59$ when
$U_S\sim\Delta$, which is much larger than previously proposed\cite{RWR}.
Introducing some bulk states in the central region with $U_S>\lambda_{\mathrm{SO}}$
was proposed to mediate inter-edge CAR\cite{RWR}, but we can see that too much bulk states at larger
$U_S$ will destroy the above subtle coherent interference
and provide more channels of transporting the electron directly to the right lead,
which contribute to the EC. During this process, there may also be bulk mediated normal reflection,
which makes $T_{\mathrm{EC}}+T_{\mathrm{CAR}}<1$.

For a visualized picture, we plot the real space charge distribution\cite{RMALY} in the device in Fig. \ref{Figchargedensity} (b),
corresponding to the dashed line in Fig. \ref{Figchargedensity} (a), with a remarkable $T_{\mathrm{CAR}}=0.59$.
Besides the central region (labeled as ``center''), some portion of the left lead (labeled as ``left'') and the right
lead (labeled as ``right'') are also included in the plot.
Since there are no remarkable bulk states involved, the charge density is mostly
distributed at the edges. However, because of coupling to the SC lead,
the electron states in the ribbon acquire an evanescent character,
and can spread over a length scale of $\xi$\cite{RWR,WC,PA}.
There are hole states at the lower edge of right lead, corresponding to the CAR channel as illustrated in Fig. \ref{Figschematic} (a).
The charge density in the upper edge
of the right lead is smaller than that in the left lead, with the difference
contributing to the CAR.
On the left lower edge, the injecting electron is perfectly
Andreev reflected, so there will be almost the same charge
density of incoming electrons and reflected holes, which
lead to almost zero charge density in total.

\begin{figure}[b]
\includegraphics[width=0.45\textwidth]{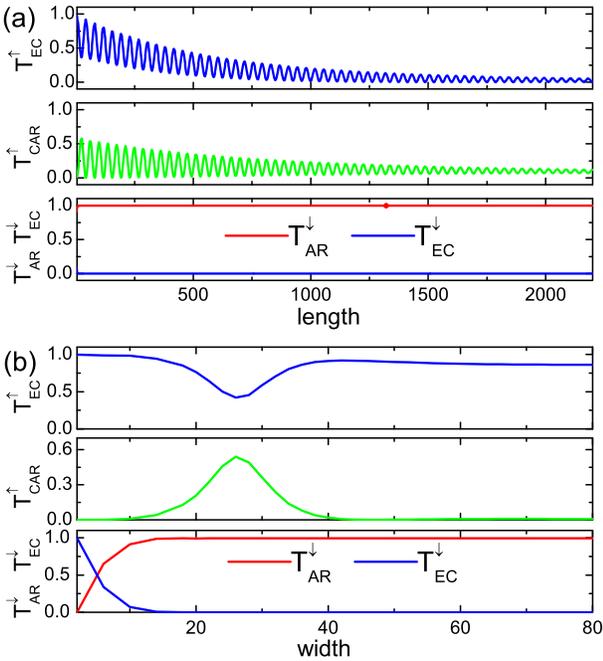}
\caption{(Color online) (a) The transmission coefficients as a
function of length with fixed width $W=26$. (b) The transmission coefficients
as a function of width with fixed length $L=20$.
Other parameters are same with
Fig. \ref{Figchargedensity}.}
\label{Figlengthwidth}
\end{figure}

Now we turn to the size effects on these transport processes.
In Fig. \ref{Figlengthwidth} (a), transmission coefficients are plotted as functions of the ribbon length.
$T_{\mathrm{CAR}}$ and $T_{\mathrm{EC}}$ oscillate rapidly with a fixed period, which can be estimated as
$\sim\pi/k_F$, with $k_F$ the Fermi wave-vector\cite{RWR}. This Fabry-Perot-type resonance\cite{WC,RWR} again reflects the
coherent interference between two origins,
the one-dimensional edge states and the inter-edge superconducting pairing via evanescent modes.
Another observation
is that the envelopes of CAR and EC of spin up channel can last for a length
much longer than the SC coherence length $\xi$, compared to previous works\cite{RWR}.
The reason should be ascribed to
the remanent robustness of the edge states, and there is almost no bulk states here \cite{YYY}. As long as there are spin up states reaching the right boundary, the CAR can occur.
However, in the limit of infinite length, both of CAR and EC decay to zero owing to the nonlocal but
evanescent nature of the quasiparticles\cite{RWR}.
On the other hand, the behavior of $T_{\mathrm{AR}}$ is simply a plateau,
since the AR is a local effect near the left boundary and has no dependence on its distance to the right boundary.

In Fig. \ref{Figlengthwidth} (b), transmission coefficients are plotted as functions of the ribbon width.
When the ribbon is extremely narrow, $T_{\mathrm{AR}}$ (red line) is not quantized until $W>16$,
due to the direct and remarkable wavefunction overlapping of states from opposite edges\cite{QFS,WC}.
Once the AR plateau is established for sufficiently large width,
it will be protected by the robust topological edge states.
For similar finite-size reasons, $T_{\mathrm{CAR}}$ (green line) is almost unobservable until $W\sim 10$.
After that, the width dependence of $T_{\mathrm{CAR}}$ is more interesting:
$T_{\mathrm{CAR}}$ has an optimal width with the largest value $\sim0.59$, and then decays to zero monotonically
in the wide limit. This vanishing of CAR is not surprising as analysed above:
The incoming electronic-like channel and the right emitting hole-like channel
are in opposite edges of the device, and this evanescent state mediated scattering will be completely
prohibited if they are sufficiently separated apart.

\begin{figure}[t]
\includegraphics[width=0.5\textwidth]{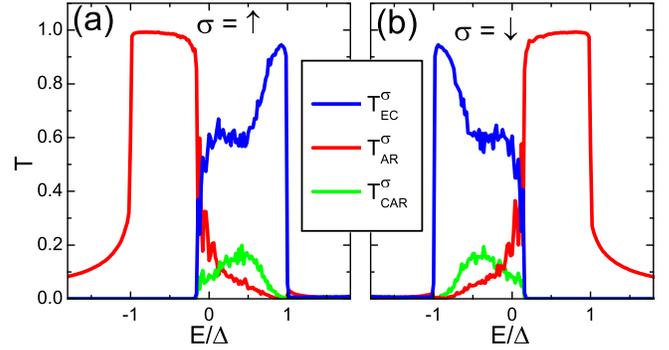}
\caption{(Color online) Disorder averaged (over 50 disorder configurations) transmission coefficients in the presence of disorder with $D=11$ meV. Other parameters are same with Fig. 2 (e) and (f) in the main text.}
\label{FigDisorder}
\end{figure}

\begin{figure}[htbp]
\includegraphics[width=0.45\textwidth]{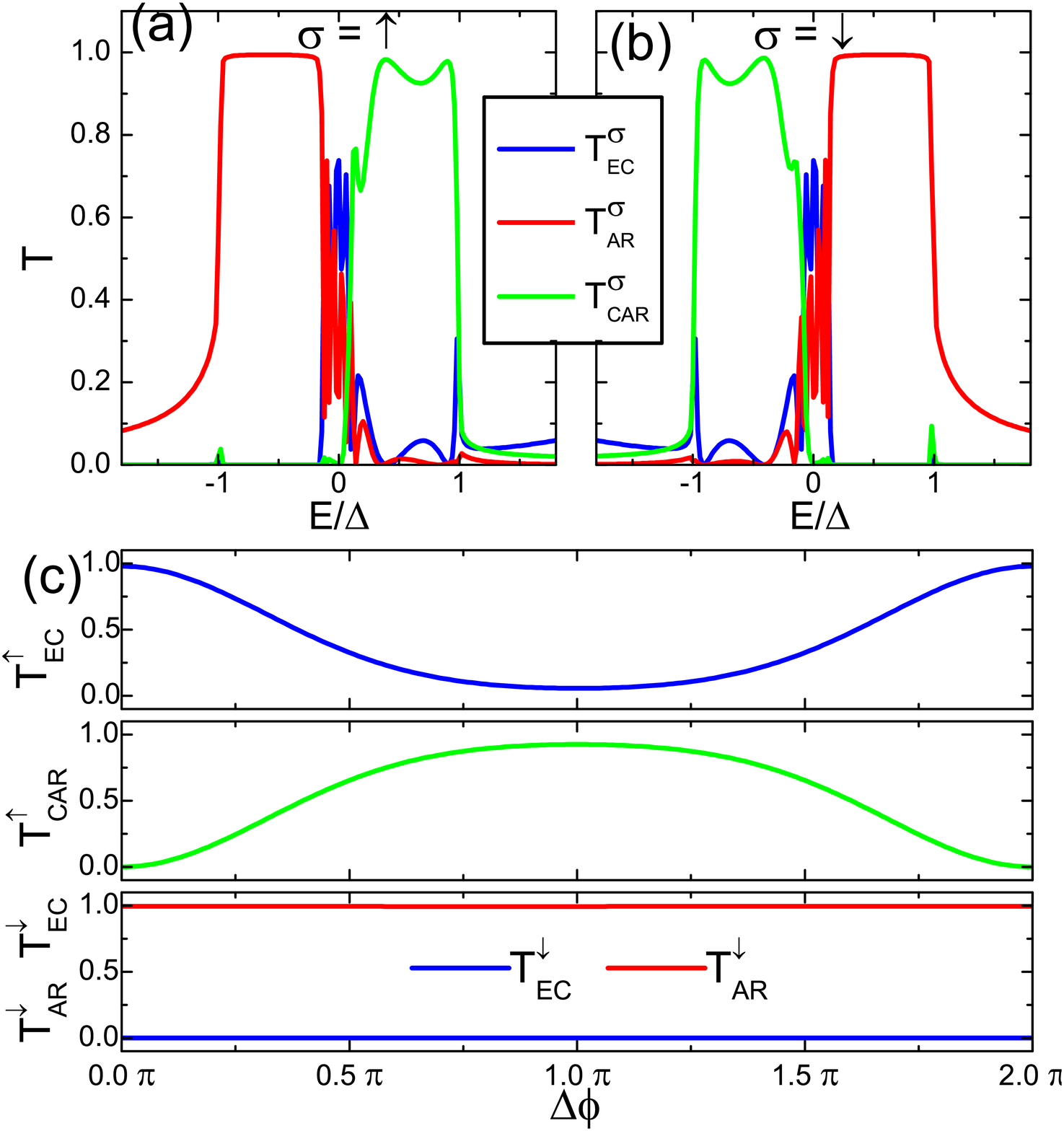}
\caption{(Color online) Transmission coefficients for the case of two edge SC leads in
Fig. \ref{Figschematic} (b). (a) The spin resolved transmission coefficients as
functions of energy with SC phase difference $\Delta\phi=\pi$ for the component of spin up (a) and spin down (b).
Other parameters are $L=40$, $W=22$, $M=2$ meV, $U_S=0$ meV.
(c) The transmission coefficients
as functions of the phase difference with energy $E=0.66$ meV.
Other parameters are same with (a) and (b).}
\label{FigPhase}
\end{figure}

\section{Disorder Effects and Josephson-Type Interference}\label{SectionDisorderJosephson}
In this section, we show two ways of inducing CAR other than nonzero $U_S$.
Instead of breaking the particle-hole symmetry from $U_S$, these two methods break
the mirror symmetry of the system. In realistic experiments, the silicene sample can hardly be perfectly mirror reflection symmetric, so the CAR should be easier to be observed.

Disorder is inevitable in realistic devices. We model this in the Hamiltonian by adding the conventional random potential to each site as
\begin{equation}
H_{\mathrm{disorder}}=\sum_{i\alpha}\epsilon_i c^{\dagger}_{i\alpha}c^{\dagger}_{i\alpha},\label{EqDisorder}
\end{equation}
where $\epsilon_i$ are independent random numbers uniformly distributed in $(-D/2,D/2)$.
Disorder break many spatial symmetries, including the reflection symmetry.
In Fig. \ref{FigDisorder}, we present the disorder averaged transmissions,
with zero $U_S$. As expected, disorder itself can also induce CAR $\sim0.2$, while leaves the AR plateau nearly untouched. We have checked that, this disorder induced CAR can persist over a wide range of disorder strength $W\sim 30$ meV, before localization happens. This also suggests that the CAR signals we discussed so far is expected to be observed in the presence of weak disorder.

Since the CAR is a result of inter-edge quantum interference,
we go one step further to investigate possible Josephson
type effects in the case of two edge SC leads with phase difference $\Delta\phi$, as illustrated in Fig. \ref{Figschematic} (b).
The calculated results are shown in Fig. \ref{FigPhase}. The prominent periodic
oscillation of $T_{\mathrm{CAR}}$ with the phase difference $\Delta\phi$ [Fig. \ref{FigPhase} (c)] confirms again that
it originates from inter-edge quantum interferences. On the other hand,
the AR, which is only a local and intra-edge phenomenon, does not show any phase dependence. We emphasise two other observations.
First, the value of $T_{\mathrm{CAR}}$ can be as large as nearly $1$.
Second, nonzero $U_S$ is now not a necessary condition for nonzero $T_{\mathrm{CAR}}$.
This is because the phase difference has broken the mirror reflection symmetry,
which is one of the crucial factors for the existence of opposite psudoparities
prohibiting CAR.

\section{The Magnitudes of SC Pairing and SOC Strength}\label{SectionPairingSOC}

\begin{figure}[htbp]
\includegraphics[width=0.5\textwidth]{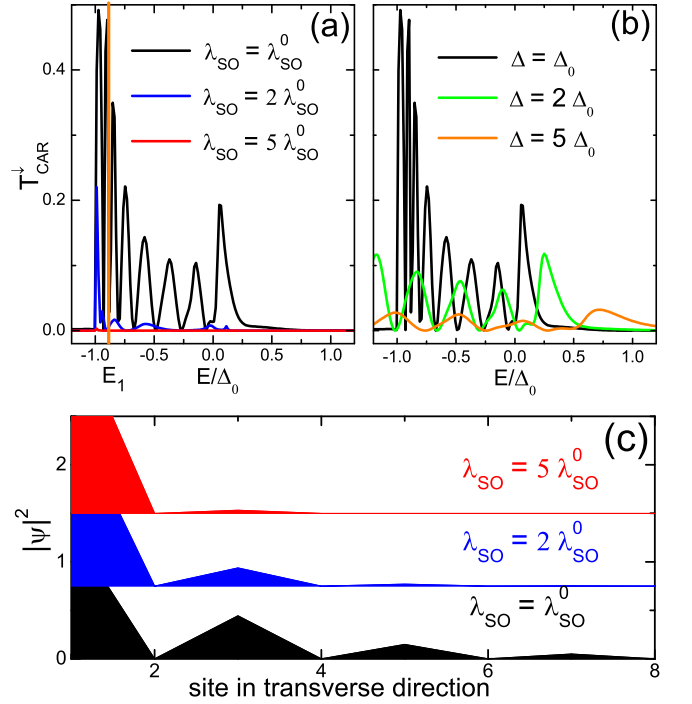}
\caption{(Color online) (a) The CAR $T_{\mathrm{CAR}}$ for different $\lambda_{\mathrm{SO}}$ .
(b) The CAR $T_{CAR}$ for different $\Delta$. (c) The distributions of wavefunctions along the transverse direction near the lower edge for
different $\lambda_{\mathrm{SO}}$ at $E_1=-0.87$ meV, corresponding to the orange vertical line in (a).
Here $\lambda_{\mathrm{SO}}^0=3.9$ meV and $\Delta_0=1$ meV, same with those taken to be constans in the main text. Other parameters are same with Fig. 2 in the main text, except that the width $W=24$.}
\label{FigLambdaDelta}
\end{figure}

As stated before, one of the key reasons for the appearance of
CAR is the similarity of magnitude between them: $\lambda_{\mathrm{SO}}\sim \Delta \sim $meV.
In Fig. \ref{FigLambdaDelta}, we plot the CARs as functions of energy, for larger values of
$\lambda_{\mathrm{SO}}$ [Fig. \ref{FigLambdaDelta} (a)] or $\Delta$ [Fig. \ref{FigLambdaDelta} (b)],
in the case of a single SC lead.
In all cases, the CAR has been remarkably suppressed.
With a larger $\lambda_{\mathrm{SO}}$ [Fig. \ref{FigLambdaDelta} (a)], the edge states are more localized in the transverse direction,
as shown in Fig. \ref{FigLambdaDelta} (c). As a result, even the evanescent modes are not capable of
transporting carriers between edges, and the CAR will be difficult to happen.
As for the case of larger $\Delta$ [Fig. \ref{FigLambdaDelta} (b)], the suppression of CAR can be attributed to thedecrease of SC coherence length for the Cooper pair forming.

\section{Conclusions}
In summary, we investigate the quantum transports through a narrow silicene-SC junction
deposits on a ferromagnet. With the help of the exchange field,
there are spin dependent AR and EC plateaus. Applying a gate voltage in the central region
can turn some fraction of the EC into CAR. Then in certain energy window
we achieve spin filtered AR, EC and CAR channels, which can be mutually spatially separated in the device.
The remarkable CAR can happen in a length range much longer than the SC coherence length $\xi$,
while with an optimal width relevant to $\xi$.
The appearance of CAR is attributed to the inter-edge coupling
through evanescent modes, determined by the close magnitude between $\lambda_{\mathrm{SO}}$
and $\Delta$. The Josephson type oscillation of CAR from two edge SC leads not only
offers larger magnitude of $T_{\mathrm{CAR}}$, but also confirms the inter-edge origin of CAR.

\section*{Acknowledgments}
We thank Yingtao Zhang, Xianxin Wu and Shu-feng Zhang for helpful
discussions. This work is supported by the Ministry of Science and Technology of China 973
program (Grant No. 2012CV821400 and No. 2010CB922904), National Science Foundation of China (Grant No. NSFC-11204294, 1190024, 11374294, 11175248, 11104339 and 61427901), and the Strategic Priority Research Program of  CAS (Grant No. XDB07000000).

\section*{Appendix A: The ``Lead Method''}
In the numerical calculations, the second method of including the superconductiviy is to couple the silicene to the semi-infinite SC lead with the Hamiltonian\cite{JW2012,JW2015,QFS,YingTaoSR}
\begin{eqnarray}
H_{\mathrm{lead}}&=&\sum_{k\alpha}(\varepsilon_{k}-\mu)b^{\dagger}_{k\alpha}b_{k\alpha}
+\sum_{k}(\Delta b^{\dagger}_{k\uparrow}b^{\dagger}_{-k\downarrow}+ \mathrm{H.c.}) \notag\\
 &+&t_{S}\sum_{\langle i,l \rangle \alpha} (c^\dagger_{i\alpha}b_{l\alpha}+ \mathrm{H. c.}),\label{EqLead}
\end{eqnarray}
where $b_{k\alpha }^{\dagger }$
is the creation operator of an electron in the SC lead with wavevector $k$ and
spin $\alpha$, and $t_S$ is the coupling strength between silicene and the SC lead. Compared to the two-terminal simulation of the ``pair method''
(because the superconductivity has been accounted into the silicene Hamiltonian in a closed form),
the ``lead method'' here is essentially a three-terminal simulation, closer to the realistic physical setup of the device.
For the convenience of numerical calculations, it is sufficient to consider the SC lead as a cluster of independent one-dimensional SC leads connecting to silicon sites\cite{RMALY,JW2012,JW2015,YingTaoSR}.

In the calculations employing Green's functions (E.g., see Appendix B), this coupling with a semi-infinite lead can be treated within the standard scheme by adding a self-energy to each site as\cite{DHL,QFSJPCM}
\begin{equation}
\Sigma^S=-ig^S\sigma_0(E+\Delta\tau_x)/2\Omega,\label{EqSCSelfEnergy}
\end{equation}
where $\tau_x$ is the Pauli matrix describing Nambu space, $g^S$ is the linewidth constant set as 2 meV, and
$\Omega=i\sqrt{\Delta^2-E^2}$ for $|E|<\Delta$ while
$\Omega=\sqrt{E^2-\Delta^2}$ for $|E|>\Delta$.

\section*{Appendix B: Transmission Coefficients in Terms of Green's Functions}
By using Green's function, the zero temperature transmission coefficients
(in units of $\frac{e^2}{h}$ throughout this manuscript) through the central region can be calculated as \cite{QFSJPCM,SGC}
$T_{\mathrm{AR}}=\mathrm{Tr}[\Gamma^L_{ee}G^r_{eh}\Gamma^L_{hh}G^a_{he}]$, $T_{\mathrm{CAR}}=\mathrm{Tr}[\Gamma^L_
{ee}G^r_{eh}\Gamma^R_{hh}G^a_{he}]$ respectively. Here $e$ ($h$) denotes the electron (hole)
degree in Nambu space. $\Gamma^{p}=i[\Sigma^r_{p}-(\Sigma^r_{p})^\dagger]$
represents the linewidth function of lead $p$ with $\Sigma^r$ denoting
the self-energy\cite{DHL}.
$G^r=[G^a]^\dagger=[E-H_C-\Sigma^L-\Sigma^R]^{-1}$
is the Green's function of the central region, where $H_C=H_{\mathrm{S}}+H_{\mathrm{pair}}$ for the ``pair method'',
and $H_C=H_{\mathrm{S}}+\Sigma^{\mathrm{S}}$ for the ``lead method'', due to their different ways of including superconductivity. Notice that all these operators, including non-SC parts, $H_C$ and $\Sigma^{L/R}$, should be expressed in the Nambu representation. The ``pair method'' simply treat the system as a two-terminal transport,
where the silicene possesses an intrinsic superconductivity.
On the other hand, the ``lead method'' simulate it as a three-terminal transport,
where the SC lead (as a bath) pins the energy region of the SC condensate.

\section*{Appendix C: BdG dispersion in the presence of the exchange field}
The bare Hamiltonian of silicene can be expressed as
\begin{equation}
H_{\mathrm{bare}}=\left(
  \begin{array}{cc}
    h^{+} & 0 \\
    0 & h^{-} \\
  \end{array}
\right),
\end{equation}
where $h^{+}$ ($h^{-}$) is the bare sub-Hamiltonian for the spin up (down) component.
In the presence of the exchange field $M$ and the SC pairing $\Delta$, the BdG Hamiltonian for the spin up (down) component is
\begin{equation}
H^{\pm}_{\mathrm{BdG}}=\left(
  \begin{array}{cc}
    h_{\pm}\mp M & \Delta \\
    \Delta & -h_{\pm}\mp M \\
  \end{array}
\right),
\end{equation}
whose eigenvalues give the location of the SC gap $(\mp M-\Delta,\mp M+\Delta)$, with $-(+)$ for the spin up (down) component.
In one word, the exchange field shifts the superconducting
gaps for opposite spins towards opposite directions, thus shifts the transmission plateaus correspondingly, as shown in Fig. \ref{FigspectrumToCAR}(a) and (b) in the main text.

\end{document}